\documentclass[12pt,letterpaper]{article}

\usepackage{subcaption}

\usepackage{setspace} % load setspace before footmisc
\usepackage{footmisc}
\usepackage{amsfonts}
\usepackage{amsmath}
\usepackage{theorem}

\usepackage[top=2.5cm, bottom=2.5cm, left=2.5cm, right=2.5cm]{geometry}

\usepackage{amsmath}
\usepackage{xcolor}

\newcommand\bovermat[2]{%
	\makebox[0pt][l]{$\smash{\overbrace{\phantom{%
					\begin{matrix}#2\end{matrix}}}^{\text{#1}}}$}#2}

\usepackage{pifont}
\usepackage{url}
\usepackage[T1]{fontenc}

\usepackage{setspace}
\usepackage{footmisc}

\usepackage[natbibapa]{apacite}
\usepackage{amssymb}
\setcitestyle{aysep={}} 

\usepackage{url}

\usepackage{mathtools}
\usepackage{amsmath}
\usepackage[latin2]{inputenc}
\usepackage{graphicx}
\usepackage{subcaption}
\usepackage{multirow}
\usepackage{esvect}
\usepackage{afterpage}

\usepackage{lipsum}
\usepackage{booktabs,array,dcolumn}

\usepackage{adjustbox}
\usepackage{graphicx}
\usepackage{pdflscape}

\usepackage{caption}
\usepackage[labelfont=bf]{caption}
\usepackage[figurename=FIGURE]{caption}
\usepackage[tablename=TABLE]{caption}

\let\ampersand\&
\renewcommand*\&{and}

\makeatletter
\def\@seccntformat#1{\@ifundefined{#1@cntformat}%
	{\csname the#1\endcsname\space}%    default
	{\csname #1@cntformat\endcsname}}%  enable individual control
\newcommand\section@cntformat{\thesection.\space}       % section-level
\newcommand\subsection@cntformat{\thesubsection.\space} % subsection-level
\makeatother

\usepackage{atbegshi}% http://ctan.org/pkg/atbegshi
\AtBeginDocument{\AtBeginShipoutNext{\AtBeginShipoutDiscard}}

\begin{document}
\setcounter{page}{1}	
%	\addtocounter{page}{-1}
\begin{titlepage}
\date{}

\begin{flushleft}
	
\title{\Large  \textbf{Direct comparison or indirect comparison via a series of counterfactual decompositions?}}

\end{flushleft}

\begin{flushleft}
	
\singlespacing{\author{Anna NASZODI}}
	%{Anna NASZODI}\\ {\textit{Research}}\\{\textit{Division,}}\\ {\textit{Sveriges Riksbank}}  \\
%	%{Anna NASZODI}\\ {\textit{Research}}\\{\textit{Department,}}\\ {\textit{Magyar Nemzeti Bank}}  \\
%	{Anna NASZODI*}\\  %{\textit{European Commission's}}  \\
%%	{\textit{Joint Research Centre}}
%	%{\textit{Via Fermi 2749 I-21027 Ispra}}\\
%	%{\textit{anna.naszodi@gmail.com}}\\
%	%{\textit{Phone:+39-0332783948}}	
%	%{\textit{anna.naszodi@ec.europa.eu}}
%	%\textit{naszodia@mnb.hu}\\ \textit{anna.naszodi@gmail.com}
%	\and
%	{Haoming LIU}\\
%	\and
%	{Jingfeng LU}\\
%	\and
%	{Francisco MENDONCA}
%  %\\  {\textit{European Commission's}}\\ {\textit{Joint Research Centre}}\\
%	%{\textit{Via Fermi 2749 I-21027 Ispra}}
%	%	{\textit{francisco.tmendonca@hotfmail.com}}
%	}	

\end{flushleft}

\thanks{Email: anna.naszodi@gmail.com.}\\

\maketitle
\setcounter{page}{1}

%\newpage
\noindent 

\singlespacing{\textbf{Abstract:}  
	%60
	We illustrate the point with an empirical analysis of assortative mating in the US, namely, that the outcome of comparing two distant groups can be sensitive to whether comparing the groups directly, or indirectly via a series of counterfactual decompositions involving the groups' comparisons to some intermediate groups. We argue that the latter approach  is  typically more fit for its purpose.}

\begin{flushleft}
\end{flushleft}

\end{titlepage}

\section{Introduction}
%\section{INTRODUCTION}

Counterfactual decompositions are commonly applied with the aim of quantifying differences between two groups. 
For instance, the aim can be to measure the degree of labour-market discrimination of an age-group (or a gender-group, or a religious-group) relative to another age-group (or another gender-group,  or another religious-group) (see \citealp{Oaxaca1973}).   
Another example is about quantifying the relative degree of marital sorting in different generations (see \citealp{NaszodiPB2019}, \citealp{NaszodiMendonca2021}).
 
There are certain applications, where the direct comparison of the two groups studied has no alternative as there is no intermediate group between them.    
E.g. there is no group between the polytheists and the monotheists that could allow us to take into account the differences between these two religious-groups gradually. 

However, there are empirical applications of counterfactual decompositions, where a series of comparisons  can serve as an alternative to the direct comparison. 
For instance, the young--old comparison on the labour-market can be conducted by comparing the young workers to the intermediate group of middle-aged workers, while also comparing the middle-aged workers to the old workers rather than comparing the two extreme age-groups directly.   

Similarly, the marital sorting of non-consecutive generations can be compared not only directly. 
Those who were young adults in 1960 %(populated mostly by the members of the early Silent Generation)
 can be compared with those who were young adults 55 years later %in 2015 
 %(populated mostly by the early Millennials)
  via the  comparisons of some  consecutive generations. % of the early Silent Generation, late Silent Generation, early Boomers, late Boomers, early GenerationX,  late  GenerationX  and early Millennials. % of the late Silent Generation, early Boomers, late Boomers, early GenerationX and late  GenerationX. 
 % While it involves the comparisons of six pairs of consecutive generations, it does not involve the direct comparison of the early Silent Generation and the early Millennials.     

In this note, we make the point that the \textit{decompositions with a series of comparisons are typically more fit for the purpose of controlling for certain effects  
than the decompositions with direct comparison}. 
This point is not new in labour economics. For instance, \cite{Richardson2013} study age-discrimination by a series of comparisons of fictitious applicants' success rates where each pair of profiles to be compared are the same along a number of traits, while differ in terms of age by a few years. 
Thereby, they control for the effect of work experience inter alia  without having to construct the counterfactuals of old workers with no work experience and young workers with 40 years of  experience.  

%compare old workers and young workers with substantially different length of work experience. 

 However, it is a novel point for the literature on assortative mating. 
Even recently, changes in the degree of sorting were commonly studied via the direct comparison of some observations distant in time. 
{For instance, 	\cite{Eika2019} in their empirical study analyzing the assortative mating--household inequality nexus  
	compare directly how American men and women were matched along the educational dimension in the years 1962 and 2013.}  
%, 1962 and 1980, 1980 and 2013.}%, as well as in some other pair of distant years.}% In particular, they construct income distributions under counterfactual scenarios where the distribution of one factor (e.g., the education level of men and women) is ďŹxed at the base year of 1962, while the other factors (including marital sorting) are measured in a year between 1962 and 2013.}        
%Eika et al distant ugyan, de tag a korcsoport, ami miatt ez nem a legjobb pĂŠlda, igaz, az is problĂŠma nĂĄluk, hogy tĂĄg a korcsoport.     
Moreover, those papers on assortative mating that conduct a series of comparisons of consecutive generations, such as   \cite{Permanyer2019}, \cite{NaszodiMendonca2021}, do not highlight  the significance of their choice. This gap is filled by our note. 

Let us see what confounding factor has to be controlled for in the context of educational assortative mating. 
Quantifying the degree of sorting is possible through its effect on a directly observed variable, the share of educationally homogmaous couples. %couples, where the education level of husbands and wives are the same. 
Changes in this share from one generation to another generation depends not only on the changes in the degree of sorting, but also on the changes in the structural availability of potential partners with various education levels. % on top of  the interaction of these factors.  
So, the  factor to be controlled for is the pair of educational distributions of marriageable men and women.

%to identify  changes in sorting through its effect, we have to net it from the effect of changing educational distributions of marriageable men and women. 
%So, changes in the education levels of marriageable men and women have to be controlled for in order to identify the changes in the degree of marital sorting. % from the changes in the share of homogamous couples across generations.      
 
Similarly to the work experience and the age of job applicants, 
the degree of sorting and the structural availability may not be independent of each other. 
For this reason, it can be difficult to construct a counterfactual generation %(or a counterfactual society),  
where the marital sorting  is the same as it was in  a certain year (e.g. 1960), while the structural availability is the same as it was in a distant year (e.g. 2015). {This difficulty is of the same source as the mental limitation preventing us to imagine a 60-years old and a 20-years old job applicant with same work experience.}   

However, it is relatively easy to construct the joint educational distributions of couples under various counterfactuals, where marital sorting and availability are measured within a reasonably short time period, e.g., a decade.    
These counterfactual distributions allow researchers to compare even very distant generations via a series of comparisons. 

In the empirical part of this paper, we apply both the direct comparison and the series of comparisons to the degree of sorting of the early Silent Generation and the early Millennials. This example illustrates that the choice is not innocuous. %\footnote{\cite{NaszodiMendonca2021} show that the outcome of some decompositions, similar to those presented in this paper, are not robust to how the counterfactuals are constructed either.} 

\section{Data and method}

We use census data on the joint educational distributions of both married and cohabiting heterosexual American couples. We refer to both types of unions as marriages. 
Similarly, we distinguish  neither between husbands and male partners, nor between wives and female partners. 

The data are from the Integrated Public Use Microdata Series (IPUMS). % from the Minnesota Population Center.
The observations are from the years  1960, 1970, 1980, 1990, 2000, 2010 and 2015. %{Some aggregated data for these  years together with the detailed results and the code of the decompositions are available on Mendeley (see \url{} ).} 
Following \cite{Eika2019}, we work with four education levels: no high school degree, high school degree, some college, tertiary level diploma.
   
In the first empirical exercise, we work with data from 1960 and 2015 covering couples where the wives are between 28 and 57 years old. 
In the second exercise, we enrich the data with observations from 1990. % while   and we restrict our analysis to marriages where the wives are between 28 and 57. 
In the third exercise, we use data from all the five intermediate decennial censuses while we restrict our analysis to marriages where the wives are between 26 and 35.  
Finally, we restrict the data  to marriages with wives between 28 and 32 for analyzing the period of 2010--2015.  
Due to these restrictions, no couple is observed twice in any of the exercises. %ne of our analyses is conducted with overlapping observations.  
Therefore, the outcome of neither of our comparisons is effected by changes in sorting over the course of individuals' lives.    
   
%In the data from 1960 and the 1970, most of the young adults observed are from the early Silent Generation and the late Silent Generation, respectively. 
%In  1980 and 1990, most of them are from the generation of the early Boomers and the late Boomers, respectively. 
%In  2000 and 2010, they are typically from the early GenerationX and the late GenerationX, respectively. 
%Finally, in 2015 we observe mostly the early Millennials as young adults.   

We apply the NM-method developed by \cite{NaszodiPB2019} and \cite{NaszodiMendonca2021} for constructing the counterfactual joint distributions. This choice is motivated by the validation exercise of \cite{NaszodiMendonca2021}: they show that while the outcomes of certain decompositions obtained with the NM are in accord with survey evidence on Americans' self-reported marital preferences, this is not the case with certain alternative methods.\footnote{Our point on the direct versus indirect comparison is robust to the choice of the method used for constructing the counterfactuals (see the Appendix).}  
 %Also, we follow them at applying the additive decomposition scheme with interaction effects developed by \cite{Biewen2014}.

%The counterfactual joint distributions allow us to  quantify the extent to which the change in marital sorting has, in itself, contributed to the change in the share of homogamous couples. 

\section{Empirical results}

\begin{figure}%[b]{0.8\linewidth}
	
	\caption{Counterfactual shares of educationally homogamous couples (in \%) -- counterfactuals are constructed by the NM} 										
	%\centering
	
	\includegraphics[width=\linewidth]{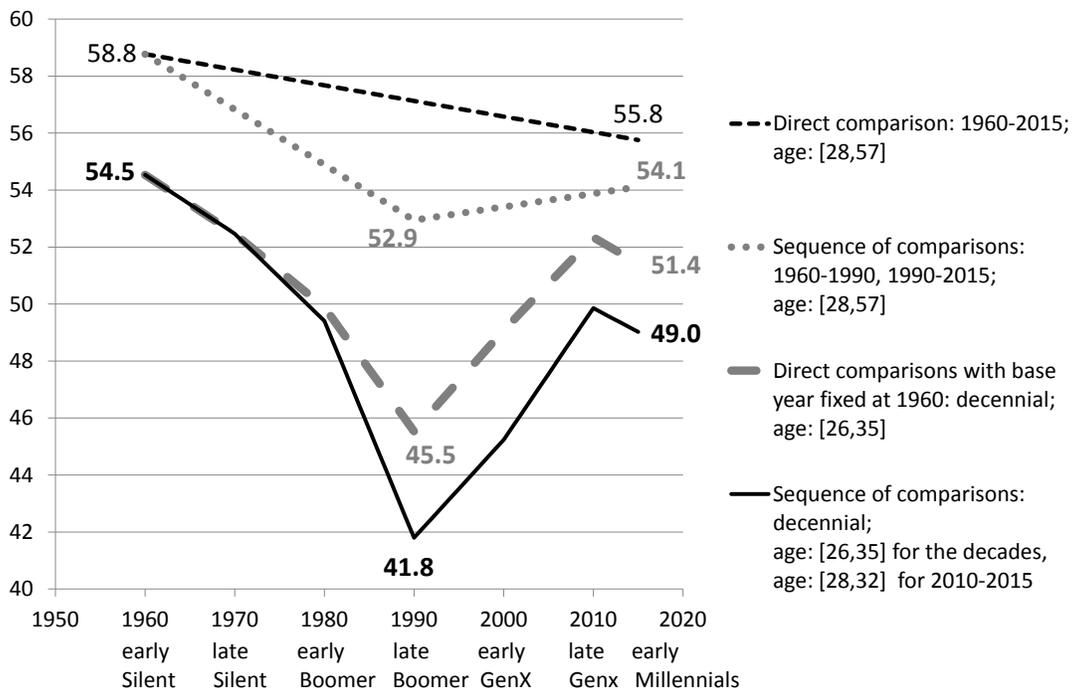}
	
	\label{fig:aggr}

\textit{Source}: author's calculations using US {census data} from IPUMS about the education level of married couples and cohabiting couples.\\ 
% in 1980, 1990, 2000 and 2010.\\
%\textit{Notes}:  The change in the share of homogamous couples attributed to the ceteris paribus  change in the degree of sorting over each decade (or the 5-years period in case of 2010--2015) is obtained by performing a decomposition  using the additive decomposition scheme with interaction effects, while the counterfactual joint distributions are constructed with the NM-method. %We use the year 1960 as a reference  year. To calculate the counterfactual shares of any other year, we add  the cumulative contributions to the observed share of homogamous couples in 1960.   
%Age of male partners is between 30 and 34 years. %Sorting along education is assumed to follow sorting along race.    
\end{figure}

Figure \ref{fig:aggr} presents  the outcomes of the decompositions. 
Its dashed black line shows the result of a direct comparison:  in 1960, 58.8\% of the observed couples with wives between 28 and 57 were homogamous. 
This share would have been decreased by  3 percentage points by 2015 provided  the education levels of young adults  remained the same as in 1960. 
If we also use the intermediate observation from 1990,  then the same effect is quantified to be higher in absolute terms (-4.7=54.1-58.8, see the gray dotted line).

If we use all the intermediate observations from the census,  
%the five census years of 1970, 1980, 1990, 2000 and 2010, 
while we also change the age group analyzed then the effect studied is found to be even higher in absolute terms  $\;\;\;$ (-5.5=49-54.5 percentage points, see the black continuous line).   
To see whether the difference is due to the shrinking age brackets, we follow \cite{Eika2019} and perform an alternative set of decompositions. 
In this exercise, we work with the [26,35] age category and keep the structural availability fixed at the base year of 1960,  while we allow the  degree of marital sorting to vary over time. % Moreover, we work with the same age category of wives (i.e., 26-35) as in our analysis of six pairs of consecutive generations. %Also, we use observations from all the seven waves.    
This alternative direct comparison-based decomposition assigns a change of the same magnitude (-3.1=51.4-54.5) to the changing sorting between 1960 and 2015 as the first direct comparison-based decomposition  (see the dashed gray line).  
%   would have been decreased by only 3.1 (=51.4-54.5) percentage points  due to the difference in the degree of sorting of young adults in the early Silent generation and the early Millennials under the counterfactual of unchanged education levels over the 55 years studied (see the dashed gray line).

%Its range is about 30\% less than that of the continuous black line.   
All in all, the changes in sorting are quantified to have contributed much more to the change in the share of educationally homogamous couples between 1960 and 2015,     
%(and thereby also to the variation in the household income inequality over 55 years) 
if their effects are measured by a series of comparisons of consecutive generations rather than by a direct comparison of non-consecutive generations.  

\section{Conclusion}

Performing counterfactual decomposition with a sequence of comparisons allows us to compare observations of two groups distant in time or in any other dimension by using intermediate observations representing gradual transition from one of the groups to the other group studied. This note promoted the gradual approach. 

We argued that this approach has the advantage relative to the direct comparison of being suitable for controlling for the effects of some confounding factors. Also, we illustrated with an empirical application that the outcome of the decomposition can be sensitive to the choice of the approach. Our example was taken from  a strand of the empirical counterfactual decomposition literature where the opportunity of sequential comparison has not always been exploited.

%First, we argued that this approach has an important advantage relative to the direct comparison of the two groups:  it is suitable for controlling for the effects of the confounding factors. Second, we illustrated with an empirical application that the outcome of the decomposition can be sensitive to the choice of the approach. Our example was taken from  a strand of the empirical counterfactual decomposition literature where the opportunity of sequential comparison has not always been exploited.

%\newpage
%\vspace{-15mm}
%\linespread{1.25}

%\bibliographystyle{apacite}
%\bibliographystyle{jpe}

\bibliography{Bib_Grad}

\newpage
\begin{center}
	
	%\textbf{Online Appendix} \\ 
	
	\textbf{Appendix} \\ 
	of the paper \\
	\textbf{Direct comparison or indirect comparison via a series of counterfactual decompositions?}\\
	
	%by\\
	%Anna Naszodi\footnote{Email: anna.naszodi@gmail.com}

\end{center}

\vspace{1cm}
	\textbf{{Appendix A: Sensitivity analysis with respect to the method used for constructing the counterfactuals}}

In this appendix, we show that our point on the direct versus indirect comparison is robust to the choice of how the counterfactuals are constructed.\footnote{We are grateful to Attila Lindner for his comment highlighting the importance of performing the related sensitivity analysis.}
 In particular, we illustrate with the same empirical application presented in the main part of the paper that the outcome of the decomposition can be sensitive to the choice between the 
 direct comparison and the sequential comparison even if the counterfactuals are constructed by the iterative proportional fitting (IPF) algorithm (rather than the NM).

The IPF algorithm, or as it is also commonly referred to, the RAS algorithm,  
is a mathematical scaling procedure which has been widely used by social scientists to standardize the marginal distribution of a contingency table  to some fixed value, while
retaining a specific association between the row and the column variables.\footnote{See \url{https://en.wikipedia.org/wiki/Iterative_proportional_fitting}.} 
The retained association is the similarity of these variables captured by the odds-ratio.

%E:\structured2016\work\papers\GradualDecomp\data\Excel
%ContT_USA_26_60_4edu_proc_2VB
%Chart2PaperGradual_IPF	
\begin{figure}%[b]{0.8\linewidth}
	
	\caption{Counterfactual shares of educationally homogamous couples (in \%) -- counterfactuals are constructed by the IPF}										
	%\centering
	
	\includegraphics[width=\linewidth]{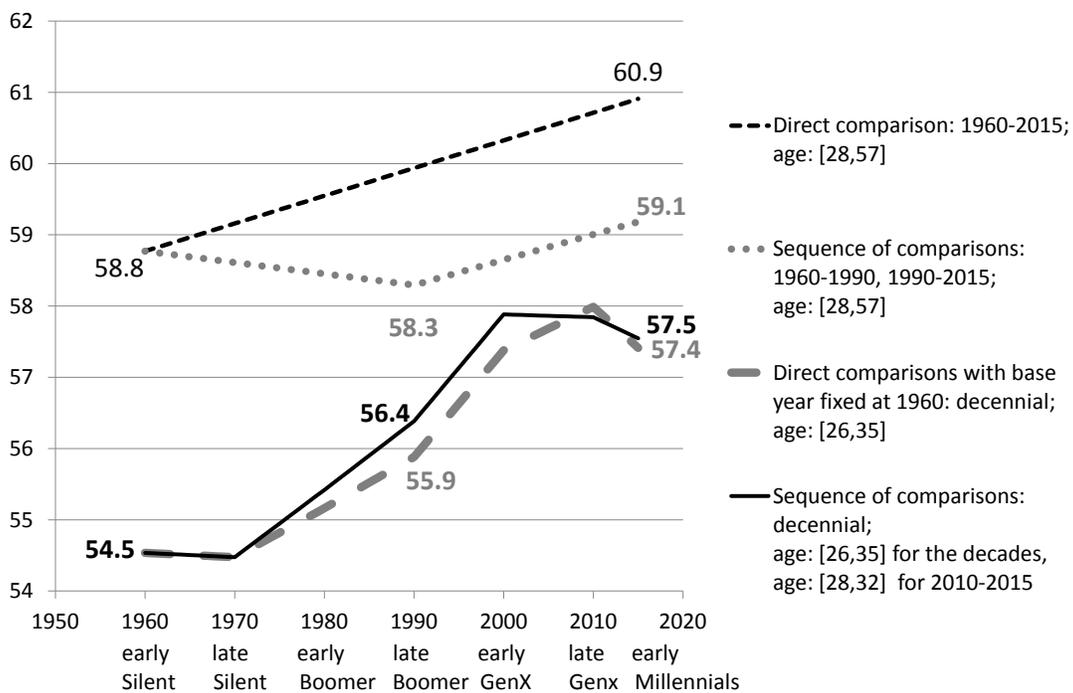}
	
	\label{fig:aggr_IPF}
	
	\textit{Source}: author's calculations using US {census data} from IPUMS about the education level of married couples and cohabiting couples.\\
	% in 1980, 1990, 2000 and 2010. 
	%\textit{Notes}:  The change in the share of homogamous couples attributed to the ceteris paribus  change in the degree of sorting over each decade (or the 5-years period in case of 2010--2015) is obtained by performing a decomposition  using the additive decomposition scheme with interaction effects, while the counterfactual joint distributions are constructed with the NM-method. %We use the year 1960 as a reference  year. To calculate the counterfactual shares of any other year, we add  the cumulative contributions to the observed share of homogamous couples in 1960.   
	%Age of male partners is between 30 and 34 years. %Sorting along education is assumed to follow sorting along race.    
\end{figure}

Figure \ref{fig:aggr_IPF} presents  the outcomes of the corresponding  decompositions. 
By comparing Figure \ref{fig:aggr} and Figure \ref{fig:aggr_IPF}, it is apparent that the identified \textit{trend of sorting is highly sensitive to how the counterfactuals are constructed}. By applying the NM, the degree of sorting along the educational displays a U-shaped trend  (see the black, continuous line in Figure \ref{fig:aggr}). As it is pointed out by \cite{Naszodi2023}, the  U-pattern is consistent not only with the forming consensus in the income and wealth inequality literature about the historical trend of the monetary dimensions of inequality, but also with the survey evidence about Americans' self-reported marital preferences.   

By contrast, the IPF-based decomposition suggests that the degree of sorting was either increasing or stagnant over the analyzed decades (see the black, continuous line in Figure \ref{fig:aggr_IPF}). This trend is similar to the trend identified by \cite{Eika2019} with their aggregate marital sorting indicator (see Figure 4 in \citealp{Eika2019}).  

Despite the monotonous increase in sorting is not corroborated by the survey evidence from the Pew Research Center analyzed by \cite{NaszodiMendonca2021}, \cite{Naszodi2023}  and  \cite{Naszodi2022_2m}, the IPF has been a popular method for constructing counterfactuals until recently.\footnote{The view is already challenged in the literature that the odds-ratio is suitable for capturing the degree of sorting (or, in general, the non-structural factor of the prevalence of homogamy) and also that the IPF is fit for constructing counterfactuals.     
\cite{NaszodiMendonca2023_RACEDU} illustrate with a numerical example that the odds-ratio violates the   
monotonicity criterion defined as the criterion against any suitable martial sorting  
measure to be monotonously decreasing in intergenerational mobility. ``The intuition behind 
the monotonicity criterion is that a society, where the pauper's son has higher chance to became the 
prince than in other societies, cannot be less open to accept marriages between paupers and 
princesses in comparison with other societies.'' 
In addition, \cite{Naszodi2023} illustrate with another numerical example that the IPF does not commute with the operation of merging neighboring categories of ordered assorted traits. This unfavorable property of the IPF allows researchers to manipulate by their choice of the categories the outcome of the decompositions performed with IPF-constructed counterfactuals, even unconsciously. Finally, \cite{Naszodi2022_2m} argues that the IPF is suitable for solving a set of problems different from constructing counterfactuals.} 
Therefore, ``old school'' researchers may find it interesting to see whether the choice between the direct comparison and the sequence of comparisons makes a difference  provided the counterfactuals can be constructed by the IPF.

Figure \ref{fig:aggr_IPF} shows that the choice matters. 
In 2015, the share of educationally homogamous couples would have been 2.1 (=60.9-58.8) percentage points higher among the couples with wives between 28 and 57 relative to 1960 under the direct comparison (see the dashed black line). Whereas the same effect is quantified to be much smaller (0.3=59.1-58.8 percentage points) if we also use the intermediate observation from 1990, but we change neither the age group studied, nor the method for constructing the counterfactuals (see the gray dotted line).

If we use all the intermediate observations from the five census years of 1970, 1980, 1990, 2000, 2010, and work with the [26,35] age category  then the same effect is measured to be  3 (=57.5-54.5) percentage points  (see the black continuous line). We obtain roughly the same effect (2.9 = 57.4-54.5) if we keep on working with the [26,35] age category, while fixing the structural availability at the base year of 1960  (see the dashed gray line).

All in all, we find that the magnitude of the effect studied can be sensitive to the choice between the series of comparisons of consecutive generations and the direct comparison of non-consecutive generations irrespective of the method used for constructing counterfactuals.\\ %However, this is not always the case. 

\textbf{{Appendix B: Counterfactual decomposition with the NM-method}}

This appendix offers a detailed explanation on how the empirical decompositions are performed with the NM-method. 
The NM-method is based on the scalar-valued sorting indicator proposed by \cite{LiuLu2006} (henceforth LL-indicator)  
and the generalized, matrix-valued LL-indicator proposed by \cite{NaszodiMendonca2021}. 
%The NM-method is based on the scalar-valued sorting indicator proposed by \cite{LiuLu2006} (henceforth LL-indicator) suitable for analyzing matching along a single dimensional dichotomous assorted trait (e.g. taking the values $L$ or $H$) and the 
%generalized, matrix-valued LL-indicator proposed by \cite{NaszodiMendonca2020} characterizing sorting along a one-dimensional {multinomial trait variable}. 
First, we introduce the original LL-indicator and the generalized LL-indicator.    
Then, we present the  NM-method. 
Finally, we introduce the decomposition scheme used. \\

%\newpage
\textbf{The original LL-indicator}\\

The LL-indicator, as it was originally developed by \cite{LiuLu2006},  is a scalar-valued, ordinal measure that can be applied 
if the \textit{assorted trait is a one-dimensional dichotomous variable} (e.g. taking the values $L$ or $H$). % and the contingency table $Z^{2\text{-by-}2}$ is a 2-by-2 matrix. 
%The LL-measure was generalized by \cite{NaszodiMendonca2020} to characterize sorting along a one-dimensional  \textit{multinomial trait variable}. 
%The generalized LL-measure is matrix-valued. First, we define the original  LL-measure, before we introduce its generalized version. 

The original LL-indicator is identical to the value taken by a function ($f:\mathbb{N}^{2\times 2} \mapsto \mathbb{R}$) that assigns a scalar to a 2-by-2 contingency table, where the contingency table is of the form 
\begin{equation}\label{Kmatrix2}
Z^{2\text{-by-}2}= \begin{bmatrix}
N_{L,L}    &  N_{L,H} \\
N_{H,L}   &  N_{H,H}
\end{bmatrix}   \;. 
\end{equation}
$N_{H,H}$ (/$N_{L,L}$) denotes the number of homogamous couples, where both spouses are $H$ (/$L$) type. 
$N_{L,H}$ (/$N_{H,L}$) stands for the number of heterogamous couples, where the husbands (/wives) are $L$-type, while the wives (/husbands) are $H$-type. 

Furthermore, we introduce the notations 
$N_{H,\cdot}=N_{H,H}+N_{H,L}$,  $N_{\cdot,H}=N_{L,H}+N_{H,H}$,  $N_{\cdot,\cdot}=N_{\cdot,H}+N_{\cdot,L}$. 
For a given triad of $\{N_{H,\cdot}, N_{\cdot,H}, N_{\cdot,\cdot}\}$,   $Q={N_{H,\cdot}N_{\cdot,H}}/ {N_{\cdot,\cdot}}$ denotes  the expected number of $H$,$H$-type couples under random matching.  We define $Q^-$ as the biggest integer 
being smaller than, or equal to, $Q$.

It is important to note that any actual realization of the joint distribution  $Z^{\text{act},{2\text{-by-}2}} \in \mathbb{N}^{2\times 2}$ with a given triad %of  $N_{H,\cdot}, N_{\cdot,H}, N_{\cdot,\cdot}$ 
can be represented by any of its cells. For instance, the actual value of the ${H,H}$ cell, i.e.,  $N^\text{act}_{H,H}$, can represent $Z^{\text{act},{2\text{-by-}2}}$, because all the other three cells' actual values are uniquely determined by the triad and $N^\text{act}_{H,H}$. 
Therefore, there is a unique ranking of the joint distributions with the same triad. 
This ranking is defined simply by the ranking of the ${H,H}$ cells: 
that table ranks higher which has higher value in its ${H,H}$ cell.   

%Therefore, for any given triad of  $\{N_{H,\cdot}, N_{\cdot,H}, N_{\cdot,\cdot}\}$, the LL-measure is defined 
%by a function $\mathbb{N} \mapsto \mathbb{R}$ that assigns a scalar value to $N^a_{H,H}$. 

The original LL-indicator defines a ranking among the joint distributions with the same, but also with different, triads by ranking their values at the ${H,H}$ cell relative to all possible values of $N_{H,H}$ conditional on the triad.  Under the assumption of non-negative sorting  (i.e., $Q^- \leq N^\text{act}_{H,H}$), the  original LL-measure is equivalent to the \textit{simplified LL-measure} defined as: 
\begin{equation}\label{LiuLusimpl0}
\text{LL}^{\text{sim}}(Z^{\text{act},2\text{-by-}2})= \frac{N^\text{act}_{H,H} -  \text{min}(N_{H,H}| N_{H,\cdot}, N_{\cdot,H}, N_{\cdot,\cdot} ) }{\text{max}(N_{H,H} |N_{H,\cdot}, N_{\cdot,H}, N_{\cdot,\cdot})-\text{min}(N_{H,H} | N_{H,\cdot}, N_{\cdot,H}, N_{\cdot,\cdot} ) }    \;. 
\end{equation}
The simplified LL-measure interprets as the ``actual minus minimum over maximum minus minimum''. %\footnote{The simplified LL-measure is almost the same as the Coleman-index defined by Equation (15) in \cite{Coleman1958}.}    
%, where 
%$\forall N_{H,H} \in\left[\text{min}(N_{H,H}|\{N_{H,\cdot},N_{\cdot,H},
%N_{\cdot,\cdot}\}),
%\text{max}(N_{H,H}|\{N_{H,\cdot},N_{\cdot,H},N_{\cdot,\cdot}\} )\right]$.

Under non-negative sorting,  $\text{min}(N_{H,H}| N_{H,\cdot}, N_{\cdot,H}, N_{\cdot,\cdot}  ) = Q^-$. And irrespective of the positive, negative, or random nature of sorting, $\text{max}(N_{H,H} | N_{H,\cdot}, N_{\cdot,H}, N_{\cdot,\cdot} )=\text{min}(N_{H,\cdot}, N_{\cdot,H} )$. By substituting these two equations to  Eq. (\ref{LiuLusimpl0}), we obtain  
\begin{equation}\label{LiuLusimpl}
\text{LL}^{\text{sim}}(Z^{\text{act},2\text{-by-}2})=  \frac{N^\text{act}_{H,H} -Q^- }{\text{min}(N_{H,\cdot}, N_{\cdot,H} )-Q^- } \;. 
\end{equation} 
Eq.(\ref{LiuLusimpl}) defines the original LL-measure under non-negative sorting, which is the empirically relevant type of sorting where the assorted trait is the eduction level.\\ 

%\newpage
\textbf{The generalized LL-indicator}\\

The first thing to note is that the LL-indicator is defined for 2-by-2 contingency tables.  
However, in the empirical part of the paper we work with a multinomial assorted trait variable as the education level can take 4 different values.  
Here, we  {relax the assumption that the assorted trait is dichotomous}.

In the \textit{multinomial case}, the one-dimensional assorted trait distribution can even be gender-specific. 
For instance, it is possible that the market distinguishes between $m\geq2$ different education levels of women and  $n \geq 2$  different education levels of men where $n$ may not be equal to $m$.   
%In addition, the joint distribution of the gender-specific assorted traits can vary over time. 
Let us denote by $Z_t$  the contingency table (of size  $n \times m$) representing the aggregate market equilibrium at time $t$.  
%(Here, we deviate from the notation used in Section \ref{sec:cc} by introducing the time index.) % in order to allow the comparison of  multiple generations.  

If both the male-specific assorted trait variable and the female-specific assorted trait variable are {one-dimensional, ordered, categorical, multinomial variables}  
then the aggregate degree of sorting at time $t$ can be characterized  by the 
\textit{matrix-valued  generalized LL-indicator} (see \citealp{NaszodiMendonca2021}).  
Its  $(i,j)$-th  element  is  
\begin{equation}\label{LiuLugengen}
\text{LL}^{\text{gen}}_{i,j} (Z_t)= 
\text{LL}( V_i  Z_t  W^T_j )    \;,
\end{equation}
where  $Z_t$ is the $n \times  m$ matrix representing the joint distribution;  
$V_i$ is the $2 \times  n$ matrix \vspace{6mm} \\ 
$V_i = \scriptsize{ \begin{bmatrix}
	\bovermat{\textit{i}}{1    & \cdots &  1} & \bovermat{\textit{n-i}}{ 0  & \cdots & 0}  \\
	0    & \cdots  & 0 & 1  & \cdots  & 1  	
	\end{bmatrix} }$   and  
$W^T_j$ is the $m \times 2$ matrix given by the transpose of \vspace{6mm} \\
$W_j = \scriptsize{ \begin{bmatrix}
	\bovermat{\textit{j}}{1    & \cdots & 1} & \bovermat{\textit{m-j}}{ 0  & \cdots  & 0}  \\
	0    & \cdots  & 0 & 1  & \cdots  & 1  	
	\end{bmatrix} }$ with   $ i \in \{1, \ldots, n-1 \} $, and  $j \in \{1, \ldots, m-1 \}$.  
This is how the LL-indicator is generalized for ordered, categorical, multinomial, one-dimensional assorted trait variables.\\ 

%The   $(i,j)$-th  element of the odds-ratio matrix is given by 
%\begin{equation}\label{oddsgen}
%\text{OR}^{\text{gen}}_{i,j} (Z_t)= 
%\text{OR}( V_i  Z_t  W^T_j )    \;. 
%\end{equation}
%%While Eq. (\ref{oddsgen}) represents $n \times m$ restrictions, Eq. (\ref{LiuLugengen}) represents $(n-1) \times (m-1)$ restrictions.\\ 

\textbf{The NM-method}\\

Next, let us see how the (generalized) LL-indicator is used by the \textit{NM-method for constructing counterfactual tables}. 
We denote the NM-transformed contingency table by  
$\text{NM}(Z_{t_p},Z_{t_a})=Z^*_{t_p, t_a}$,  
where the degree of sorting is measured at time $t_p$, while availability is measured at time $t_a$.
Unlike  $Z_{t_p}$ and $Z_{t_a}$,  $Z^*_{t_p, t_a}$  cannot be observed. %So, it needs to be constructed from the former two tables.  

%Apparently, here we deviate from the notation used in Section \ref{sec:cc}. 
In the empirical examples presented in the paper,  $t_p$ corresponds to the year when a relatively old generation is observed. 
Moreover,   $Z_{t_p}$ corresponds to table $K$ representing the joint educational distribution of couples in this old generation.  
Also,  $t_a$ corresponds to the year when the educational distribution of marriageable men and women in a relatively younger generation is observed. 
Finally,  $Z^*_{t_p, t_a}$ corresponds to table $K^{\text{yg}}$ representing the counterfactual joint educational distribution of couples in the younger generation.

The counterfactual table $Z^*_{t_p, t_a}$ should meet the following two conditions:   
$\text{LL}^{\text{gen}}(Z^*_{t_p, t_a})= \text{LL}^{\text{gen}}(Z_{t_p})$ in order to make the aggregate degree of sorting the same under the counterfactual as at time $t_p$.   
While the condition on availability is given by a pair of restrictions of    
$Z^*_{t_p, t_a}  e^T_{m}=  Z_{t_a}  e^T_{m}$ and  $ e_{n} Z^*_{t_p, t_a}= e_{n} Z_{t_a}$, where $e_{m}$ and  $e_{n}$ are all-ones row vectors of size $m$ and $n$, respectively.    

First, we present the solution for $Z^*_{t_p, t_a}$  in the simplest case, where the 
assorted trait variable  is dichotomous, %(i.e., that can take  the values $L$ and $H$ only),  
before we introduce the solution for the multinomial case. %an ordered, categorical, multinomial, one-dimensional assorted trait. 
In the \textit{dichotomous case}, the counterfactual table $Z^*_{t_p, t_a}$ to be determined is a 2-by-2 table, just like the observed tables  
$Z_{t_p}=\begin{bmatrix}
N^p_{L,L}    &  N^p_{L,H} \\
N^p_{H,L}   &  N^p_{H,H}
\end{bmatrix}$ and 
$Z_{t_a}=\begin{bmatrix}
N^a_{L,L}    &  N^a_{L,H} \\
N^a_{H,L}   &  N^a_{H,H}
\end{bmatrix} $.  
The solution for its cell corresponding to the number of  ${H,H}$-type couples is: 
\small
\begin{equation}\label{Solution}
N^*_{H,H}   =
\frac{\left[  N^p_{H,H} - \text{int}\left(\frac{N^p_{H,\cdot}N^p_{\cdot,H}} {N^p}\right)\right]  \left[{\text{min}\left(N^a_{H,\cdot}, N^a_{\cdot,H} \right)- \text{int}\left(\frac{N^a_{H,\cdot}N^a_{\cdot,H}} {N^a} \right) }\right]     }{\text{min}\left(N^p_{H,\cdot}, N^p_{\cdot,H} \right)- \text{int}\left(\frac{N^p_{H,\cdot}N^p_{\cdot,H}} {N^p} \right) }     
+\text{int}\left(\frac{N^a_{H,\cdot}N^a_{\cdot,H}} {N^a}\right)   \;, 
\end{equation}
\normalsize 
where $N^p_{H,H}$ is the number of ${H,H}$-type 
couples observed at time $t_p$. Similarly,  
$N^p_{H,\cdot}$ (the number of couples, where the husband is $H$-type),   $N^p_{\cdot,H}$ (the number of couples, where the wife is $H$-type), and  ${N^p}$ (the total number of couples) are also observed at time $t_p$.\footnote{For the derivation of Eq. (\ref{Solution}) and for the proof of uniqueness of this solution, see  \cite{NaszodiMendonca2021}.}   
While $N^a_{H,\cdot}$, $N^a_{\cdot,H}$, and ${N^a}$ are observed at time  $t_a$. 
So, Equation (\ref{Solution}) expresses $N^*_{H,H}$ as a function of variables with known values.  
Regarding the values of all the other three cells of $Z^*_{t_p, t_a}$, those can be calculated from $N^*_{H,H}$  by using the condition on the row totals and column totals of $Z^*_{t_p, t_a}$.

Next, let us see how the  NM-method works in the \textit{multinomial} case, where the counterfactual table  $\text{NM}(Z_{t_p},Z_{t_a})=Z^*_{t_p, t_a}$, as well as $Z_{t_p}$ and $Z_{t_a}$ are of size $n \times m$.   
It is worth to note that $\text{NM}(Z_{t_p},Z_{t_a})$ depends on the row totals and column totals of $Z_{t_a}$, but not on  $Z_{t_a}$ itself.  
So, instead of thinking of the NM-method as a function mapping $\mathbb{N}^{n \times m} \times \mathbb{N}^{n \times m} \mapsto \mathbb{R}^{n \times m}$, we should rather think of it as a function mapping $\mathbb{N}^{n \times m} \times \mathbb{N}^{n} \times \mathbb{N}^{m} \mapsto \mathbb{R}^{n \times m}$. Accordingly, we will use the following alternative notation:   
$\text{NM}(Z_{t_p}, Z_{t_a}  e^T_{m}, e_{n} Z_{t_a})$  as well. 

With this new notation,    
%Next, let us see how the NM-method works, if the contingency table is of size $n \times m$.   
the problem for the \textit{multinomial, one-dimensional assortative trait} can be formalized as follows.  
Our goal is to determine the transformed contingency table $Z^*_{t_p, t_a}$ of size $n \times m$ under the restrictions 
given by the target row totals and the target column totals observed at time ${t_a}$:  $R_{t_a}:= Z_{t_a} e^T_{m}= Z^*_{t_p, t_a} e^T_{m}$, and  $C_{t_a}:= e_{n} Z_{t_a}=e_{n} Z^*_{t_p, t_a}$.   
%, where $e_{m}$ and  $e_{n}$ are all-ones row vectors of size $m$ and $n$, respectively.  
The additional restriction is  $\text{LL}^{\text{gen}}(Z^*_{t_p, t_a})=\text{LL}^{\text{gen}}(Z_{t_p})$.

By using Eq.(\ref{LiuLugengen}), we can rewrite the problem  as follows.   
We look  for $Z^*_{t_p, t_a}$, where  
$V_i  R_{t_a} = V_i  Z^*_{t_p, t_a} e^T_{m} $, and  $C_{t_a}  W^T_j = e_{n} Z^*_{t_p, t_a} W^T_j $;  and 
$\text{LL}(V_i  Z_{t_p}  W^T_j)= \text{LL}(V_i  Z^*_{t_p, t_a}  W^T_j)$ 
for all   $i \in \{1,..., n-1\} $ and $j \in \{1,..., m-1\} $. The matrices  
$V_k$ and $W_p$ are defined the same as under Eq.(\ref{LiuLugengen}).   
For each $(i,j)$-pairs, these equations define a problem of the 2-by-2 form.       
Each problem  can be solved separately by applying Eq.(\ref{Solution}).      
The solutions determine $(n-1) \times (m-1)$ entries of the $Z^*_{t_p, t_a}$ table. 
The remaining $m+n-1$ elements of the $Z^*_{t_p, t_a}$ table can be determined with the help of the target row totals and target column totals. \\

%\textbf{The IPF-algorithm}\\
%
%The IPF algorithm applied to tables $Z_{t_p}$ and $Z_{t_a}$ is defined by the following two steps to be iterated until convergence.  
%First, it factors the rows of the seed table $Z_{t_a}$ in order to match the row totals of $Z_{t_p}$.  
%The table obtained after the first step (to be denoted by $Z^{'}_{t_a}$) may not have its column totals equal to the column totals of $Z_{t_p}$. 
%In this case, it is necessary to perform a second step.         
%
%As the second step, the IPF factors the columns of $Z^{'}_{t_a}$ to match the corresponding column totals of $Z_{t_p}$. 
%The table obtained  after  this  step (to be denoted by $Z^{''}_{t_a}$) may not have its row totals equal to the row  totals of $Z_{t_p}$. 
%In this case, repeating the first step is necessary with $Z_{t_a}=Z^{''}_{t_a}$. 
%Alternatively, we stop the iteration. % $Q^{IPF}=Q^{''}$. % and call $Q^{''}$ the IPF-transformed table.  
%%These two steps are repeated until convergence, i.e., until the table remains practically unchanged after a new step. 
%The table constructed by the IPF is the last value of $Z^{''}_{t_a}$.\\   

\textbf{Decomposition scheme}\\

As to the decomposition scheme, we apply the additive scheme with interaction effects proposed  by \cite{Biewen2014}.
For two factors ($A_{t_a}$ and $P_{t_p}$) and two time periods ($0$ and $1$), it is  %with interaction effect is given by 
\begin{multline}\label{eq:Bdecom2}
f(A_1, P_1)-f(A_0, P_0)  = 
\overbrace{[f(A_1, P_0)-f(A_0, P_0)]}^{\mbox{\text{{\normalsize due to }}} \Delta \mbox{\textit{\normalsize A}}}+  
\overbrace{[f(A_0, P_1)-f(A_0, P_0)]}^{\mbox{\text{{\normalsize due to }}} \Delta \mbox{\textit{\normalsize  P}}}\\
+\underbrace{[f(A_1, P_1)-f(A_1, P_0) - f(A_0, P_1)+ f(A_0, P_0)]}_{\mbox{\text{\normalsize due to the joint effect of }} \Delta \mbox{\textit{\normalsize  A}} \mbox{\text{\normalsize { and }}}  \Delta \mbox{\textit{\normalsize  P}}}  \;,
\end{multline}
where function $f(A_{t_a}, P_{t_p})$ maps the space spanned by the two factors  into  $\mathbb{R}$. 

In the empirical analysis,  $f(A_{t_a}, P_{t_p})$ determines the share of educationally homogamous couples in a population of a generation, where  the structural  availability is the same as in  $A_{t_a}$, while the aggregate degree of sorting is the same as in $P_{t_p}$. Function $f$ is the composition of function $h$ and $g$ as $f(A_{t_a}, P_{t_p})=h\circ g(A_{t_a}, P_{t_p})$, where $g(A_{t_a}, P_{t_p})$ constructs the counterfactual contingency table  if ${t_a} \neq{t_p}$, otherwise it is equal to $Z_{t_p}$. The  counterfactuals constructed for our empirical analysis involving the comparison of consecutive generations are $g(A_{1960}, P_{1970})$, $g(A_{1970}, P_{1960})$, 
$g(A_{1970}, P_{1980})$, $g(A_{1980}, P_{1970})$,
$g(A_{1980}, P_{1990})$, $g(A_{1990}, P_{1980})$, 
$g(A_{1990}, P_{2000})$, $g(A_{2000}, P_{1990})$, 
$g(A_{2000}, P_{2010})$, $g(A_{2010}, P_{2000})$, 
$g(A_{2010}, P_{2015})$, $g(A_{2015}, P_{2010})$.  
While $h(Z)=  \sum_{i=1}^{n} \sum_{j=1 | j= i}^{m} Z_{i,j} / \sum_{i=1}^{n} \sum_{j=1}^{m} Z_{i,j}$ determines the ratio of the sum of the diagonal cells to the sum of all the cells of table $Z$.  
The  counterfactual table $g(A_{t_a}, P_{t_p})$ for ${t_a} \neq{t_p}$  is  $\text{NM}(Z_{t_p}, Z_{t_a}  e^T_{m}, e_{n} Z_{t_a})$.  

%Depending on the choice between the NM and the IPF, the  counterfactual table $g(A_{t_a}, P_{t_p})$ for ${t_a} \neq{t_p}$  is either  $Z^{\text{NM}}_{t_p, t_a}=\text{NM}(Z_{t_p},Z_{t_a})$, or, 
%$Z^{\text{IPF}}_{t_p, t_a}=\text{IPF}(Z_{t_p},Z_{t_a})$. %The choice has to be made by the researcher.

%Under the assumption that the search and matching mechanism is frictionless, this share is the function of   
%(i) the observed availability $A_t$, i.e., the educational distributions of marriageable men and women at time $t$; 
%(ii)  the directly unobservable preferences over the partners' education level  $P_t$; and 
%(iii) the interaction of availability and preferences. 

%In Eq. (\ref{eq:Bdecom2}), $f(A_1, P_0)$, and $f(A_0, P_1)$ represent the shares of educationally heterogamous couples under the  \textit{counterfactuals} that the factors are measured at different points in time.  For instance, $f(A_1, P_0)$ is the share of educationally heterogamous couples in an imaginary generation whose gender-specific educational distributions are identical to the gender-specific educational distributions of the generation most active on the marriage market at $t=1$ and whose marital  educational preferences are identical to the marital educational preferences of the generation most active on the marriage market at $t=0$. 
%These counterfactuals are constructed with the NM-method.     

\end{document}